\documentclass{PoS}

\title{Finite Size Scaling and Universality in SU(2) at Finite Temperature}

\ShortTitle{Finite Size Scaling and Universality in SU(2) at Finite Temperature}

\author{A. Denbleyker$^1$, \speaker{Yuzhi Liu}$^{1, 2}$, Y. Meurice$^1$, and A. Velytsky$^3$ \\
        \llap{$^1$}Department of Physics and Astronomy, The University of Iowa,
Iowa City, Iowa 52242, USA\\
				\llap{$^2$}Kavli Institute for Theoretical Physics China, CAS, Beijing 100190, China\\
				\llap{$^3$}Physics Department, Brookhaven National Laboratory, Upton NY 11973, USA\\
				E-mail: \email{alan-denbleyker@uiowa.edu}\\
        E-mail: \email{yuzhi-liu@uiowa.edu}\\
        E-mail:\email{yannick-meurice@uiowa.edu}\\
        E-mail:\email{vel@quark.phy.bnl.gov}}

\abstract{
We study the 4-th Binder cumulant on $4\times{N_\sigma}^3$ lattices for a pure SU(2)
gauge theory. We use 20 data points for a sequence of ${N_\sigma}$  in $\beta$ intervals
shrinking when ${N_\sigma}$ increases, in order to reduce the nonlinear effects. 
Using a log-log fit of the slope versus ${N_\sigma}$, we obtain the preliminary result 
$\nu=0.637(11)$ in reasonably good agreement with the value for the $3D$ Ising model universality class. The corrections due to irrelevant directions 
appear to be dominated by a term proportional to ${N_\sigma}^{-2.03(4)}$ which seems compatible with the breaking of rotational symmetry. 
}

\FullConference{The XXVII International Symposium on Lattice Field Theory - LAT2009\\
		 July 26-31 2009\\
		 Peking University, Beijing, China}

\begin{document}

\section{Introduction}
Finite Size Scaling (FSS) provides a powerful tool which is used to extrapolate information to infinite volume. It can also be used to study the critical behavior and calculate critical exponents. 
When the system is close to a 2nd order phase transition, some long distance properties can be determined by the global features of the system, such as the symmetries and the dimensionality. 
Universality enables us to classify seemingly different systems into certain classes.  
Svetitsky and Yaffe  \cite{svetitsky} pointed out that, for $d+1$  dimensional $SU(N)$ pure gauge model, ``if any portion of the boundary is second-order, then the critical behavior will be described by some fixed point of $d$-dimensional, $Z(N)$ invariant spin systems.''
For a pure gauge $SU(2)$ theory in 3+1 dimension,  we expect the universality class of the finite temperature transition to be the same as 
the 3 dimensional Ising model. Existing results on FSS for $SU(2)$ \cite{engels89,fingberg92,bogo,papa,avbinder} agree well with this expectation.

In the following, we work on $SU(2)$ gauge theory in 3+1 dimensions: $N_\tau\times N_\sigma^3$, where $N_\tau = 4$, and $N_\sigma = 2, 4, 6, 8, \cdots 16$, with periodic boundary conditions. We focus on direct  
estimations of the exponents $\nu$ and $\omega$. For $N_\tau = 4$, the only direct estimate of $\nu$ we are aware is $0.65(4)$
obtained in Ref. \cite{engels89}, which is compatible with the
accurate value $0.6298(5)$ obtained in Ref. \cite{hasenbusch} 
for the 3-dimensional Ising model.  We would like: 1) to improve the accuracy of the estimate of Ref.  \cite{engels89} and 2) resolve the corrections due to the 
irrelevant directions. Part 2) is largely unexplored and a better understanding of these corrections could help us design methods to 
reduce these effects as done in Ref. \cite{hasenbusch} for the 3 dimensional Ising model. 
In  the existing work for $N_\tau = 4$ in Refs. \cite{engels89,fingberg92,avbinder}, a fixed $\beta$ interval procedure was used.  This means the $\beta$ interval is fixed for different volumes. In these proceedings, we shrink the interval in order to reduce the nonlinear effects \cite{ymbinder} and use a finer $\beta$ resolution. 
\section{Binder cumulants and FSS}
In the following, we define the 4th order Binder 
cumulant \cite{binder}, $g_4$, as
\begin{equation}
	g_4=1-\frac{\langle P^4\rangle}{3\langle P^2\rangle^2}, 
	\quad P=\frac1{N_\sigma^3}\sum_{\vec{x}}\frac12{\rm Tr}\prod_{\tau=1}^{N_\tau}U_{\tau,\vec{x};0}
	\end{equation} 
Related definitions appear in the literature, such as $B_4=\frac{\langle P^4\rangle}{\langle P^2\rangle^2}$.
We assume that there is no external field and that the $g_4$ depends on the scaling variables as
\begin{equation}
	g_4 = g_4(u_{\kappa} N_\sigma^{1/\nu},u_1 N_\sigma^{-\omega},\dots) 
	\label{eq:g4}
	\end{equation}
with 
	\begin{eqnarray*}
	u_{\kappa}&=&\kappa+u_{\kappa}^{(2)}\kappa^2+\dots\\
	u_1&=&u_1^{(0)}+u_1^{(1)}\kappa +\dots
	\end{eqnarray*}
where $u_{\kappa}$ is the only relevant scaling variable,  $u_1$ is the first irrelevant scaling variable, and $\kappa$ is the reduced quantity $\kappa = (\beta-\beta_c)/\beta_c$.
Expanding up to the first nonlinear corrections, we obtain	
\begin{equation}
	g_4(\beta,N_\sigma) = g_4(\beta_c,\infty) +f_1\kappa N_\sigma^{1/\nu}+f_2\kappa^2 N_\sigma^{2/\nu}+(c_0+c_1\kappa N_\sigma^{1/\nu}) N_\sigma^{-\omega}+\cdots
	\label{eq:g4exp}
	\end{equation}
This expansion is accurate if $|\kappa| N_\sigma ^{1/\nu}$ is small enough. In addition we would like the nonlinear effects of $f_2$ and 
$c_1$ to be negligible. It is easy to estimate $f_2$ from numerical data for intermediate values of $\kappa$ \cite{ymbinder} . However, $c_1$ 
is more difficult to resolve from an already small effect and its effect will be ignored. In the following, we will work with values of $|\kappa| N_\sigma ^{1/\nu}$ such that the effects of $f_2$ are within the numerical errors of $g_4$  that we now proceed to discuss. 
\section{Error Analysis}
For each volume, we worked on the region near $\beta_c$ (we discuss $\beta_c$ later). We generated 50,000 Polyakov loops with 20 different seeds for each $\beta$. Therefore, for each $\beta$, we have 1,000,000 values. However, the  data is correlated and we will need to remove the correlations. 
\begin{figure}[h]
  \begin{tabular}{l@{\hspace*{20mm}}r}
    \includegraphics[width=0.45\textwidth]{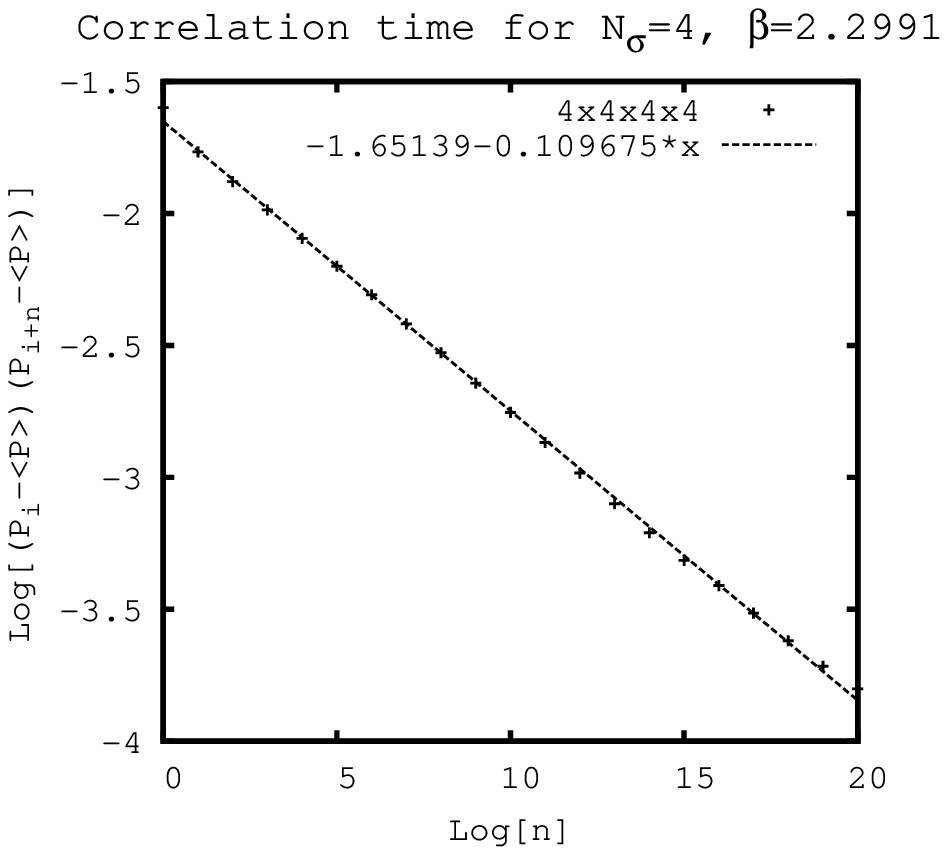} &
    \includegraphics[width=0.45\textwidth]{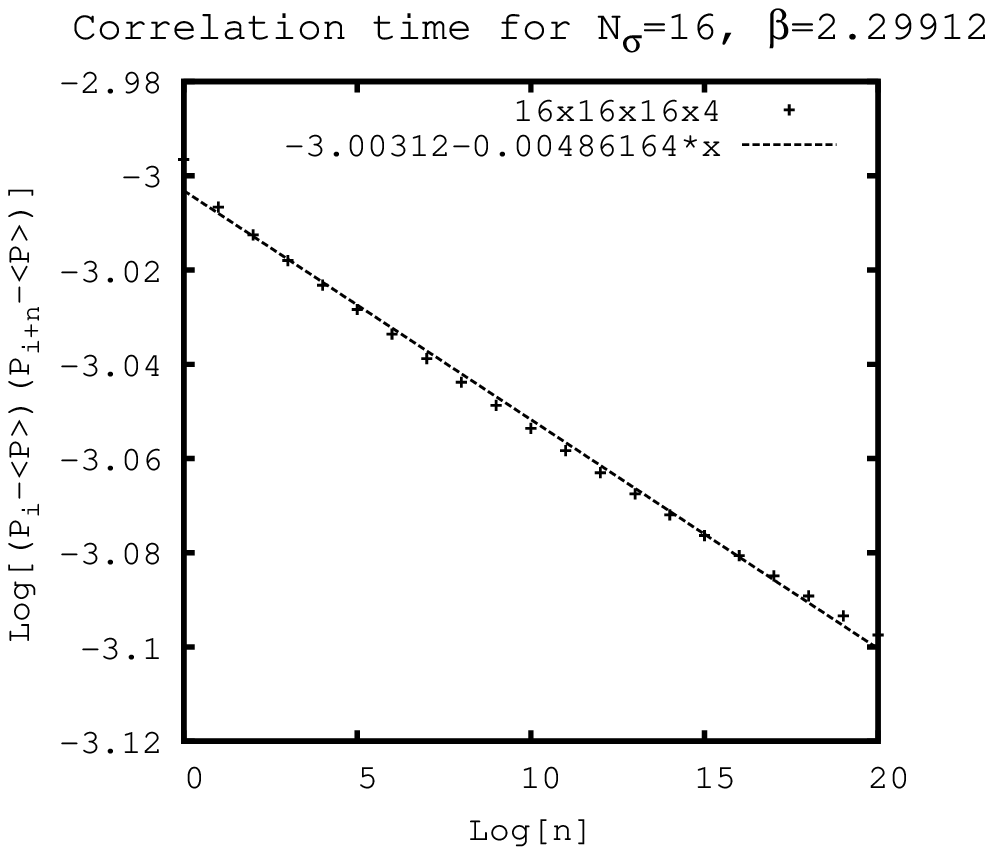} \\
  \end{tabular}
  \caption{ These two graphs show that the autocorrelation time changes with the volume. For $N_\sigma$=4, the autocorrelation time is around 10. For $N_\sigma$=16, the autocorrelation time is around 200 illustrating the critical slowing down. }
  \label{fig:cortime}
   \begin{center}
 \includegraphics[width=0.55\textwidth]{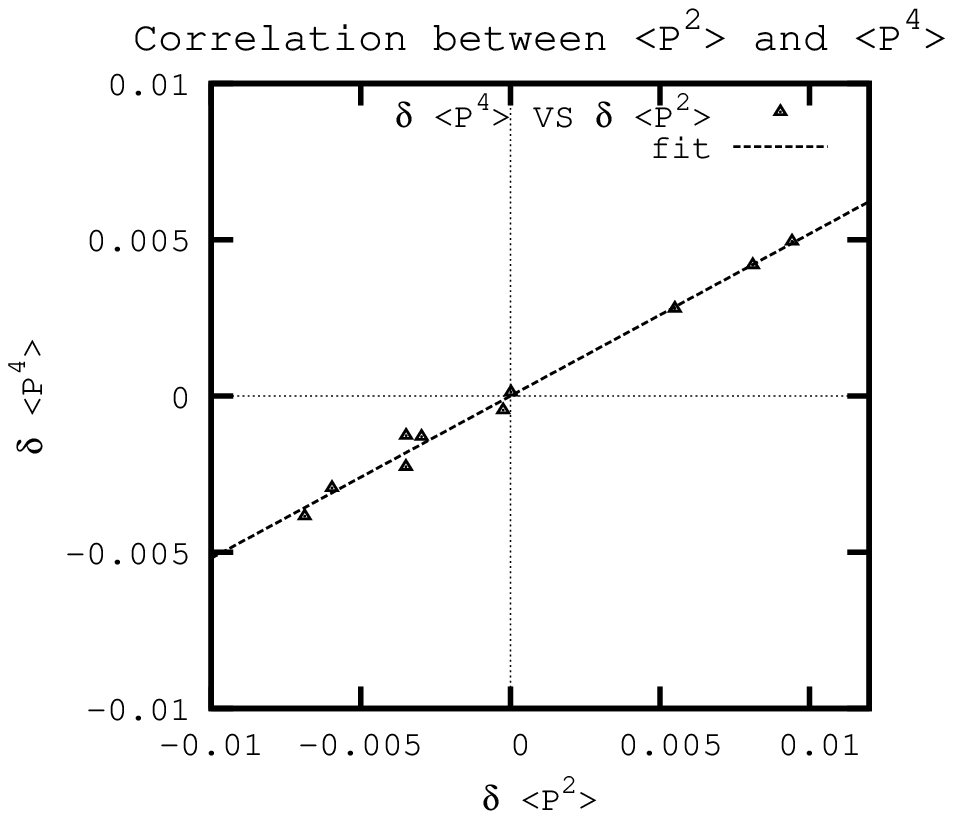} 
  \caption{variation of $P^4$ and $P^2$ from the ensemble average of for 10 different seeds . }
  \label{fig:err}
  \end{center}
 \end{figure}
The autocorrelation times change with $\beta$ and the volume. Fig. \ref{fig:cortime} shows two autocorrelation functions.  In order to remove the autocorrelation among the data, we tried the bootstrap and jackknife methods. For the jackknife method, we skipped every $\tau$ data in order to remove the autocorrelation, where $\tau$ is the autocorrelation time. Comparisons with other methods to estimate the 
errors will be discussed elsewhere.
We should note here that one should not calculate the error of $P^4$ and $P^2$ first and then use the error propagation to get the error of $g_4$. This is simply because $P^4$ and $P^2$ are highly correlated, see Fig. \ref{fig:err}. The jackknife method can remove the correlation between $P^4$ and $P^2$. 
 \section{Determination of the critical exponent $\nu$}
\begin{figure}
  \begin{tabular}{l@{\hspace*{20mm}}r}
    \includegraphics[width=0.45\textwidth]{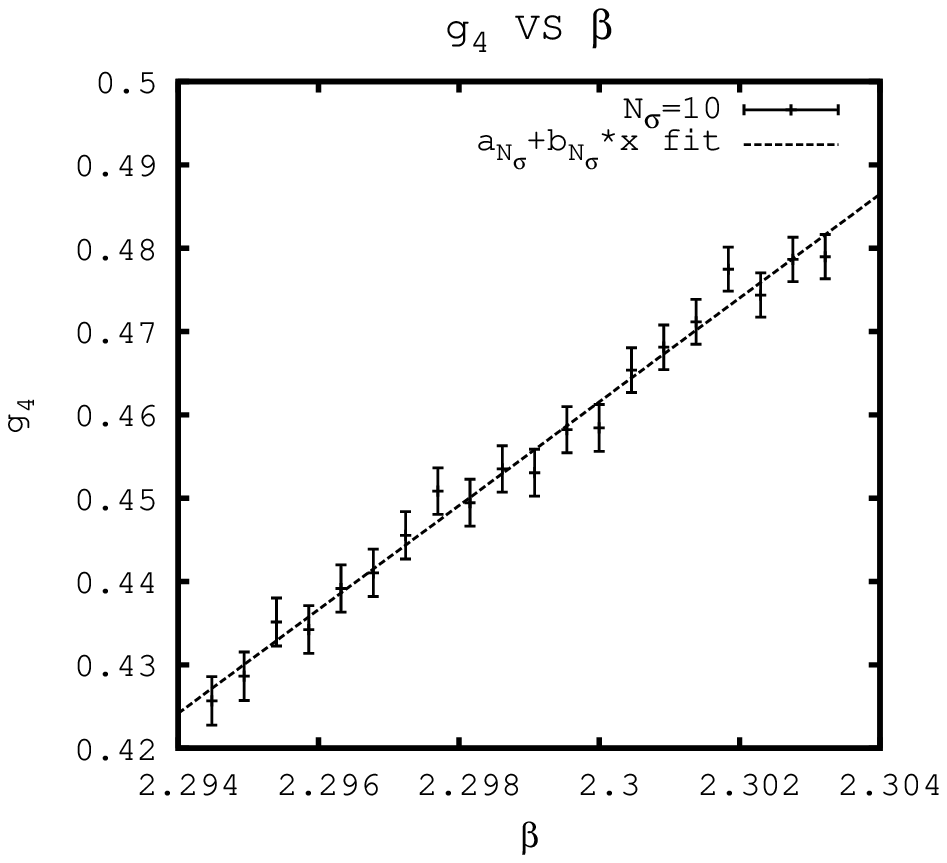} &
    \includegraphics[width=0.45\textwidth]{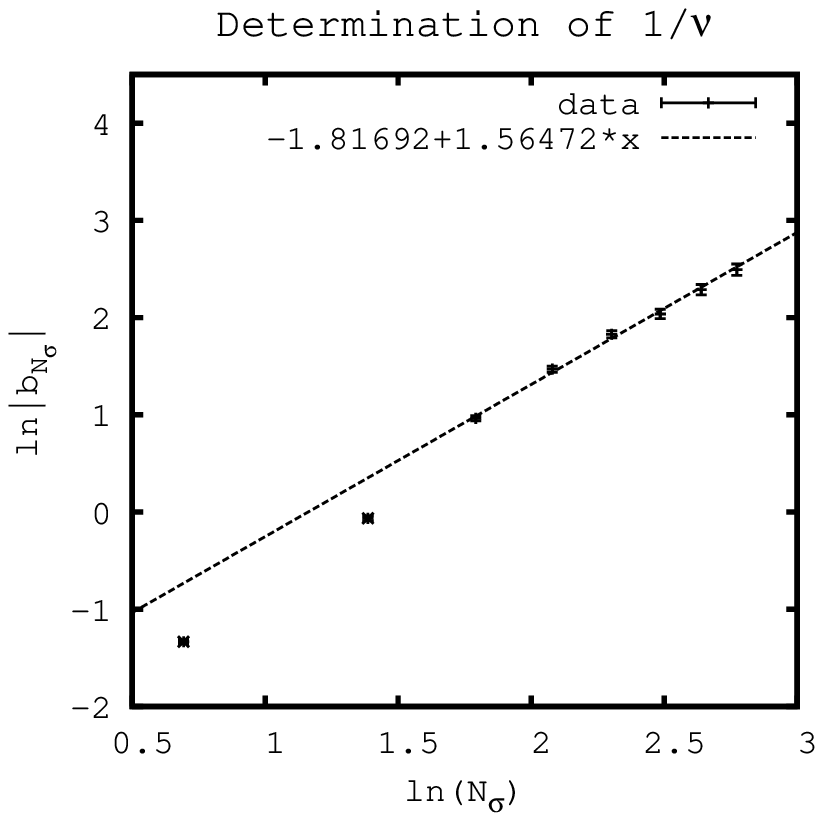} \\
  \end{tabular}
  \caption{Left:  linear fit of $g_4$ near $\beta_c$ for $N_\sigma = 10$, $({\beta_c})_{app} = 2.299$ and ${(1/\nu)}_{app} = 1.6$. We get $a_{N_\sigma}$ and $b_{N_\sigma}$ from the fit for different volumes.
  Right: determination of $1/\nu$ from the log-log fit discussed in the text. $b_L\simeq f_1\times N_\sigma^{1/\nu}/\beta_c$. By taking $Log[|b_{{N_\sigma}}|]$ vs. $Log[N_\sigma]$, we can get the slope which is just $1/\nu$. We only did this fit with $N_\sigma$ between 6 and 16. The $N_\sigma$ = 2, and 4 points deviate from the linear approximation. }
  \label{fig:g_betalogb_logN}
\end{figure}

\begin{figure}
  \begin{tabular}{l@{\hspace*{20mm}}r}
    \includegraphics[width=0.45\textwidth]{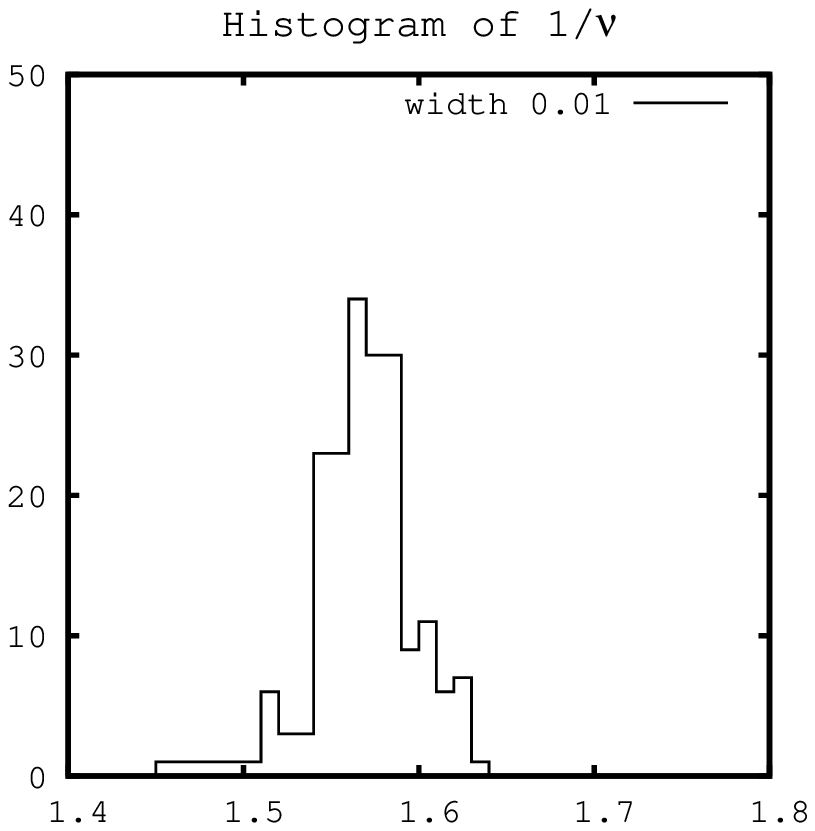} &
    \includegraphics[width=0.45\textwidth]{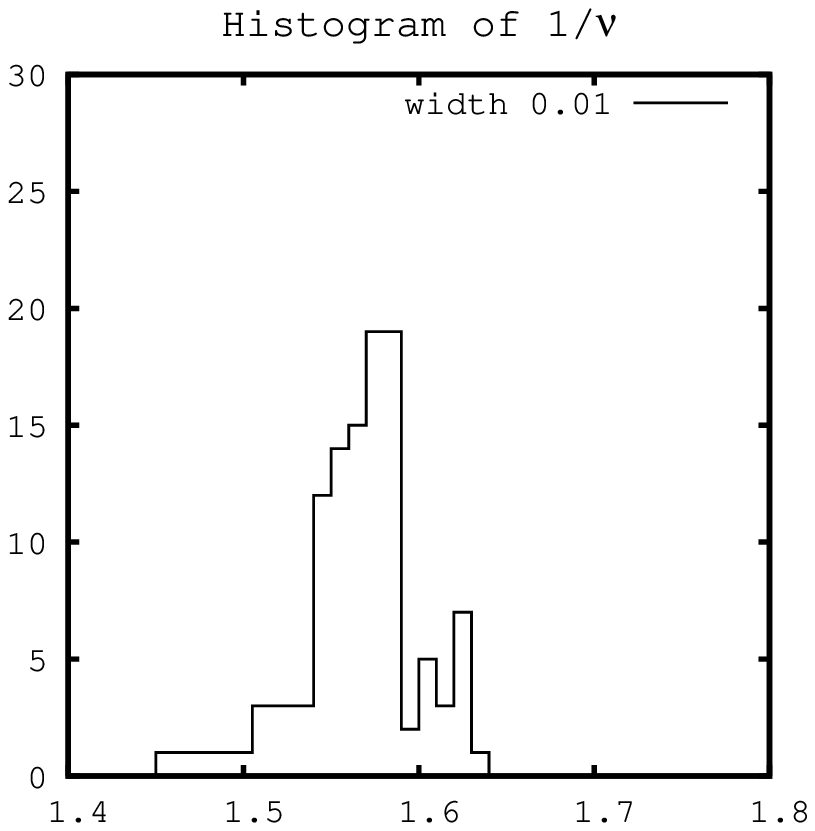} \\
  \end{tabular}
  \caption{ Left Fig.: $({\beta_c})_{app}$ changes from 2.297 to 2.301; $({1/\nu})_{app}$ changes from 1.4 to 1.8. $\overline{1/\nu}=1.570$; $\sigma=0.027$.
 Right Fig.: $({\beta_c})_{app}$ changes from 2.298 to 2.300; $({1/\nu})_{app}$ changes from 1.4 to 1.8. $\overline{1/\nu}=1.571$; $\sigma=0.028$.}
  \label{fig:histo}
\end{figure}

We now focus on the estimation of $\nu$. 
If we are reasonably close to $\beta_c$, we 
can use the linear form: 
\begin{equation}
 g_4(\beta,N_\sigma) \simeq g_4(\beta_c,\infty) +c_0 N_\sigma^{-\omega} +f_1\kappa N_\sigma^{1/\nu}
 \label{eq:linear}
 \end{equation}
This expression contains 6 unknown parameters: $g_4(\beta_c,\infty), \ c_0, \  \omega  , \  f_1,\  \beta_c$ and ${1/\nu}$ and we will use a new strategy to attack this difficult problem \cite{ymbinder2}. A first observation is that the dependence on $\nu$ can be isolated from the other parameters by 
studying the linear dependence in $\beta$. However, one should keep in mind that the slope of this linear fit depends implicitly on the choice of the center of the interval which should be as close as possible to $\beta_c$. 
In order to guarantee that nonlinear effects are under control, we will consider  $\beta$ intervals of the form
 \begin{equation}
  |\beta - ({\beta_c})_{app}|<0.015\times (4/N_\sigma)^{({1/\nu})_{app}} 
  \end{equation}
We will start with reasonable values for $({\beta_c})_{app}$ and ${(1/\nu)}_{app}$  and then show that  the effect of their variations is small.
The factor 0.015 has been chosen following a procedure described in Ref.   \cite{ymbinder} and guarantees 
an approximate linear relation between 
 $g_4$ and $\beta$  at fixed $N_{\sigma}$:
\begin{equation}
g_4 \simeq a_{N_\sigma}+b_{N_\sigma}\times \beta \ .
\end{equation}
This is illustrated in the left side of Fig.  \ref{fig:g_betalogb_logN} for $({\beta_c})_{app} = 2.299$, ${(1/\nu)}_{app} = 1.6$ and $N_\sigma =10$. We see that in the chosen interval the deviation from linearity seem mostly due to statistical errors rather than a systematic 
curvature that would be observed if we had chosen a broader interval.
The volume dependence of the parameters of the linear fit are 
\begin{eqnarray}
	\label{eq:a_linear}
	&&a_{N_\sigma} \simeq g_4(\beta_c,\infty) +c_0 N_\sigma^{-\omega} -f_1 N_\sigma^{1/\nu}\\
	&&b_{N_\sigma} \simeq f_1 N_\sigma^{1/\nu}/\beta_c
	\label{eq:b_linear}
	\end{eqnarray}
Once we have $b_{N_\sigma}$ for different $N_\sigma$, we can do a log-log fit to determine the inverse critical exponent $1/\nu$. 
One should note that the slope can be determined independently of 
$\beta_c$ or $g_4(\beta_c,\infty)$. 
This is illustrated in the right side of  Fig.  \ref{fig:g_betalogb_logN}.
From the log-log fit with $N_\sigma \geq 6$, we obtain $1/\nu=1.56(4)$. We plan to model and explain the deviations from linearity 
at low $N_\sigma$. 
We need to address the dependence on ${(\beta_c)}_{app}$ and ${(1/\nu)}_{app}$.  We used the central value $\beta_c=2.2991$ from Ref. \cite{avbinder} as the critical value and changed the center of the interval $({\beta_c})_{app}$,  between 2.297 and 2.301 and  $({1/\nu})_{app}$ between 1.4 and 1.8. We calculated  $1/\nu$ for a  
set of 189 values of $({(\beta_c)}_{app}, \ {(1/\nu)}_{app})$. The histogram is shown in Figure \ref{fig:histo} left. We can see that the values appear in a rather large range between 1.468 and 1.631. The average of this set of results is 
$\overline{1/\nu}=1.570$ with $\sigma =0.027$ which will be our preliminary result. At the conference, a larger value was found, but 
with low statistics at large $N_\sigma$. In the meantime, we collected more data at large $N_\sigma$. 
If we choose the center of the data in a smaller range, namely with ${(\beta_c)}_{app}$ from 2.298 and 2.300 and the same range for ${(1/\nu)}_{app}$ , we get a slightly different distribution shown on the right side of \ref{fig:histo}, but the average is essentially the same. 
It seems thus possible to average over ${(\beta_c)}_{app}$ and ${(1/\nu)}_{app}$ in order to get a more accurate value of $1/\nu$. 
For reference, 
${1/\nu}_{Ising}$ is estimated as 1.5887(85) in Ref. \cite{landau} and 
1.5878(12) in Ref. \cite{hasenbusch} and our preliminary result here is  consistent with these more accurate values. This is preliminary, we plan to go to larger $N_\sigma$ and to determine  $\beta_c$ independently using the method of Ref. \cite{avbinder}. 
\section{Determination of the critical exponent $\omega$}
\begin{figure}
\begin{center}
      \includegraphics[width=0.75\textwidth]{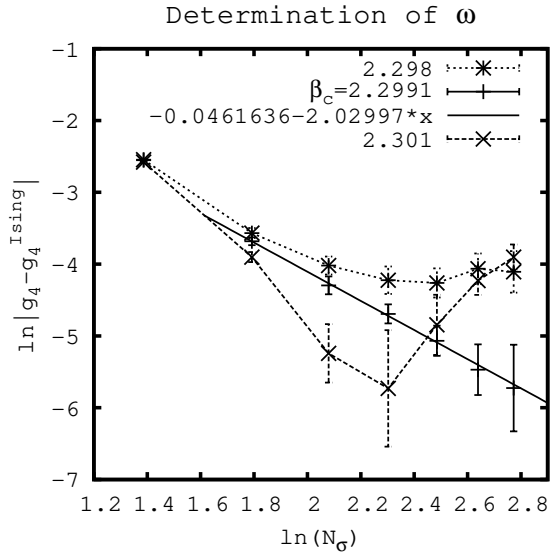}   \caption{For $\beta_c$=2.2991, the behavior is approximately  linear: $g_4\simeq  g_4(\beta_c,\infty) +c_0\times N_\sigma^{-\omega}$. }
  \label{fig:omeg}
  \end{center}
 \end{figure}

Unless we determine $\beta_c$ and $1/\nu$ very precisely, it is very difficult to subtract the effects of the third term of  Eq. (\ref{eq:linear}).
If we can work at $\beta_c$,  this term is absent:
\begin{equation}
g_4(\beta_c,N_\sigma) =g_4(\beta_c,\infty) + c_0 N_\sigma^{-\omega}
\end{equation}
Consistently with the previous section and the rest of the literature, we assume the universal value $g_4(\beta_c,\infty)=0.46575$ as found in Ref. \cite{hasenbusch}. 
$Log[|g_4-g_4(\beta_c,\infty)|]$ vs. $Log[N_\sigma]$ should be linear right at $\beta_c$ and nonlinear for all the other $\beta$s. 
This is shown in Figure \ref{fig:omeg}. At the same time, the slope is $-\omega$. The result we obtained from this analysis is $\omega=2.030(36)$. 
This is very different from $\omega_{Ising}=0.812$ \cite{landau}. It is possible that the coefficient of the $N_\sigma^{-\omega}$ is very small and the exponent we extrapolated is a sub-subleading exponent. For a detail discussion of the subleading corrections, see Ref. \cite{hasenbusch}. The most plausible explanation seems that this exponent is related to the irrelevant direction associated with the 
breaking of rotational symmetry \cite{campo} and which is close to 2.

\section{Conclusion and perspective}

By using  methods discussed in Refs. \cite{ymbinder,ymbinder2}, we analyzed the 4-th order Binder cumulant of pure SU(2) lattice gauge theory and estimated the critical exponent $\nu$ and $\omega$. $1/\nu$ is in good agreement with the value for the universality class of the 
$3D$ Ising model, while the corrections to scaling seem dominated by anisotropic effects \cite{campo}. 
\begin{acknowledgments}
This research was supported in part  by the Department of Energy under Contract No. FG02-91ER40664. Y. Liu thanks the KITPC for its hospitality while this work was in progress and the participants of the  2009 Les Houches Summer School for valuable discussion and comments. 

\end{acknowledgments}

\end{document}